# Knowledge of Process-Structure-Property Relationships to Engineer Better Heat Treatments for Laser Powder Bed Fusion Additive Manufactured Inconel 718

Thomas G. Gallmeyer[1], Senthamilaruvi Moorthy[1], Branden B. Kappes[1], Michael J. Mills[2]

Behnam Amin-Ahmadi[1], and Aaron P. Stebner[1]

[1]Alliance for the Development of Additive Processing Technologies (ADAPT), Colorado School of Mines, Golden, CO 80401, USA

[2]Materials Science and Engineering, Ohio State University, Columbus, OH 43212, USA

**Abstract**

Dislocation structures, chemical segregation, $\gamma'$, $\gamma''$, $\delta$ precipitates and Laves phase were quantified within the microstructures of Inconel 718 (IN718) produced by laser powder bed fusion additive manufacturing (AM) and subjected to standard, direct aging, and modified multi-step heat treatments. Additionally, heat-treated samples still attached to the build plates vs. those removed were also documented for a standard heat treatment. The effects of the different resulting microstructures on room temperature strengths and elongations to failure is revealed. Knowledge derived from these process-structure-property relationships was used to engineer a super-solvus solution anneal at 1020 ºC for 15 minutes, followed by aging at 720 ºC for 24 hours heat treatment for AM-IN718 that eliminates Laves and $\delta$ phases, preserves AM-specific dislocation cells that are shown to be stabilized by MC carbide particles, and precipitates dense $\gamma'$ and $\gamma''$ nanoparticle populations. This "optimized for AM-IN718 heat treatment" results in superior properties relative to wrought/additively manufactured, then industry standard heat treated IN718: relative increases of 7/10% in yield strength, 2/7% in ultimate strength, and 23/57% in elongation to failure are realized, respectively, regardless of as-built vs. machined surface finishes.

# 1. Introduction

## 1.1 Motivation

Recent works have established that dislocation cells that form as a result of rapid solidification during laser powder bed fusion (L-PBF) (also called "selective laser melting" (SLM)) additive manufacturing (AM) may impart higher strengths and elongations to failure upon 316L stainless steel and CoCrFeNiMn high entropy alloys than the best established traditional manufacturing methods are able to achieve [1–3]. Because these cells are orders of magnitude smaller than the grain structure of the alloys, the strength of alloys may be augmented through a microstructure-based size effect and/or dislocation hardening; a modified Hall-Petch relationship with a contribution from dislocation cells has shown to model the strength enhancements [3]. Furthermore, because the cells "modulate" dislocation motion and promote nano-twin formation during mechanical deformation, as opposed to blocking dislocation motion like high angle grain boundaries (GBs), this strength augmentation may be realized without the strength-ductility tradeoff that usually limits other strengthening mechanism available to traditionally manufactured alloys [2]. Altogether, this knowledge has established a means to engineer AM alloys to perform beyond the strength-ductility tradeoff limits established for traditionally manufactured alloys.

Still, the aforementioned stainless steel and high entropy alloys are predominantly single-phase alloys that do not require sophisticated post-processing heat treatments; they do contain carbides and oxides, but those particles are induced during processing and largely unaltered by heat treatments [1–3]. It is logical that dislocation cell networks should also benefit the mechanical performances of any alloy system that exhibited such substructures because of L-PBF AM. However, understanding of the process-structure-property relationships in alloys that are typically strengthened by multiple phases, especially how the complicated, multi-step heat treatments required to achieve those phase transformations after AM may affect the desired dislocation substructures, is incomplete.

Nickel superalloys is a class of such alloys: they are known to exhibit dislocation cells due to L-PBF processing, and those cells are known to impact mechanical properties [4–6]. In this work, sufficient understanding of the process-structure-property relationships of L-PBF AM Inconel 718 (IN718) subjected to post-processing heat treatments is established to enable purposeful design of better heat treatments for AM-IN718 materials. Specifically, dislocation structures, chemical segregation, $\gamma'$, $\gamma''$, $\delta$ precipitates and Laves phases are concurrently quantified and systematically evaluated. It is demonstrated that this understanding can be used to engineer post-AM heat treatments that result in AM-IN718 materials that exhibit better strengths and elongations to failure than industry-standard wrought IN718 materials – heat treatments that are better than those being recommended by manufacturers and adopted by the AM industry today.

In today's commercial practices, standard heat treatments (SHTs) developed for traditionally manufactured (wrought, cast, forged, powder metallurgy, etc.) IN718 are predominantly used and recommended for AM-IN718 [7–10]. These SHTs consist of solution annealing and aging steps [11] to facilitate microstructure homogenization and precipitation strengthening, respectively. More specifically, high temperature solution annealing of AM-IN718 can homogenize elements that segregate during AM, dissolve eutectic phases back into the matrix [7,12,13], and promote recovery of high dislocation densities produced during printing [4,14]. Subsequent multi-step aging treatments after solution annealing shows improved yield strength (YS) and ultimate tensile strength (UTS) over the as-printed condition, comparable to the same effects these aging steps have on traditionally manufactured IN718 after solution treatments [13,15–17]; namely, the strengthening is attributed to dense $\gamma'$ and $\gamma''$ nanoprecipitate structures.

The intuition behind recommending these same heat treatments for AM-IN718 was that the initial solution anneal step of the SHT would "normalize" the microstructures of the AM-IN718 material – it was anticipated that after the solution anneal step, AM-IN718 would be essentially the same as traditionally processed IN718 at this same stage of heat treatment. Then, the subsequent heat treatment steps would have exactly the same effects (which are reviewed in further detail in Section 1.2), and the end result would be AM-IN718 materials that meet the same specifications as traditional manufactured IN718 materials. However, it is now well established that AM-IN718 materials are not the same after the solution anneal step in the heat treatment [12,13]. In using usual solution annealing temperatures more than 900 ºC but less than 1100 ºC, AM-specific dislocation cells do not recover; rather, carbide and oxide particles provide stability to their structures, as well as grain boundaries [12,13]. To fully "reset" the microstructures of AM-IN718, recrystallization at temperatures above 1100 ºC is necessary. However, grain growth, in addition to oxide and carbide particle growth at GBs also occur during recrystallization [12,13], which are well known to diminish mechanical performances [18,19].

Recently, Pröbstle et al. [4] reported superior creep performance from multiple heat treatment conditions of L-PBF AM-IN718 relative to cast and wrought IN718, which reached a maximum when a solution anneal temperature of 1000 ºC was used in the aforementioned SHT process vs. 930 °C. They credited the AM-specific dislocation cells for better creep performance at low strains, but hypothesized that the best overall creep performance arose from the absence of δ phase allowing for higher volume fractions of $\gamma'$ and $\gamma''$ precipitates when choosing the anneal temperature to be greater (1000 ºC) vs. lower (930 °C) than the δ-solvus temperature (~ 1010 ºC) [4]. However, in traditionally manufactured IN718, δ phase is necessary to pin GBs, preventing grain coarsening during creep. While Pröbstle et al. noted that δ phase was not needed to control grain sizes of AM-IN718, the mechanistic reason is yet to be fully explained.

More recently, Sui et al. [16] reported upon the room temperature strength-elongation tradeoff of AM-IN718 made by a powder-fed laser directed energy deposition (DED) process as a function of varying heat treatment parameters. They showed that changing heat treatments could result in higher yield strengths (~ 1290 MPa)[1], ultimate strengths (~ 1535 MPa)[1], and elongations to failure (19.9%)[1]. They primarily focused on characterizing different Laves phase structures and their effects on monotonic tensile properties after being directly aged vs. first being solution annealed at 1050 ºC for 15 vs. 45 min prior to the same aging treatment. The authors proposed that small and granular Laves phase particles improved yield strength by impacting the γ″ phase volume fraction, size, and precipitate distribution. However, while subgrain structures appear to be similar in the as-manufactured samples via low magnification micrographs, closer examination shows that the subgrain structures are not cellular, but rather their morphology is dominated by eutectic formations - a primary difference of the DED AM-IN718 materials relative to what has been reported for L-PBF AM-IN718 [4,12,20–22]. This difference provides strong evidence that the process-structure-property relationships for L-PBF AM-IN718 materials likely differ than for the DED material.

Furthermore, several microstructure features known to be important to attain the best performances from traditionally manufactured IN718 are yet to be studied in AM-IN718, such as

---

[1] Engineering stress & strain values from their paper have been converted to true stress & strain values to allow direct comparison to the analyses in this paper.

γ′/γ″ coprecipitation [23]. Complete knowledge of the interactions of all of the microstructure features of AM-IN718 is necessary to develop and qualify better heat treatments for AM-IN718 materials with confidence and purpose. This work begins by first understanding the initial, as-printed microstructure in conjunction with each step of one SHT recommended for L-PBF AM-IN718 today, including the structure-property relationships of: 1) as-printed, 2) solution annealed at 980 ºC for 1 h, and 3) complete SHT conditions. We aim to holistically understand what is different about AM-IN718 after the SHT relative to cast and wrought IN718 and why. We also consider two variations relative to applications of these heat treatments in industry: heat treating the entire build at once, with parts still attached to the build plate, vs. heat treating individual parts already removed from the build plate; surprising 100 MPa differences in yield strengths between these two conditions are found to result from substantially different microstructures. We then study the effects of single-step heat treatments at temperatures below solution anneal temperatures on the AM-IN718 microstructures and mechanical properties to better understand the individual roles of the microstructure constituents. Finally, we demonstrate that the holistic process-structure-property relationship knowledge can be used to engineer better heat treatments for L-PBF AM-IN718.

**1.2 Additional knowledge of additively vs. traditionally manufactured Inconel 718 process-structure-property relationships**

Before presenting the new work, knowledge of IN718 structure-property relationships and differences already known in AM-IN718 are reviewed, especially for readers not familiar with this specific nickel superalloy.

IN718 is a Ni-based superalloy used in aerospace applications, especially in gas turbines, due to its mechanical stability at operating temperatures up to 650 °C [24–26]. Much of its high temperature strength originates from two intermetallic phases that are precipitated through aging treatments: γ′ ($Ni_3(Al,Ti)$, ordered face-centered cubic (fcc)) and γ″ ($Ni_3Nb$, ordered body-centered tetragonal (bct)) [24,27–29]. While γ′ is known as a strengthening precipitate in other Ni-base superalloys, it plays a minor role in IN718 strengthening due to low coherency strains it creates within the γ fcc matrix (less than ~1.25%) [30]. Instead, γ″ serves as the main strengthening precipitate since it produces up to 2.9% coherency strain within the matrix [30].

In addition to their monolithic morphologies, γ′ and γ″ can also form coprecipitate structures in various configurations [31]. The γ′ unit cell size is in between that of the γ matrix and γ″ precipitates. Therefore, it is proposed that γ′ - matrix interfaces serve as preferential nucleation sites for γ″ [28], because the lattice strain energy caused by γ″ nucleating on γ′ - matrix interfaces is lower than that of γ″ nucleating completely within the matrix [23,32,33]. Coprecipitates have been shown to improve mechanical responses by requiring more complex dislocation structures to induce plastic deformation [34,35]. They also enhance thermal stability as their coarsening rates at elevated temperatures are lower than the monolithic precipitates due to a combination of interface-controlled kinetics (i.e. reducing total elastic energy through interfaces [23]) and diffusion-controlled kinetics (i.e. reduced solute flux across coprecipitate interfacial layers [31]). However, the γ′/γ″ coprecipitation found to be optimal for traditionally manufactured IN718 materials has yet to be reported for AM-IN718

The control of precipitate coarsening in IN718 is critical since γ″ tends to transform to the equilibrium phase, δ ($Ni_3Nb$, ordered orthorhombic), through nucleation on existing stacking faults within γ″ precipitates [36] at temperatures between 700 °C and 1000 °C [27,28]. Unlike γ″, the δ phase is incoherent with the matrix, resulting in limited impact on the precipitate strengthening of the alloy [29], δ phase does inhibit grain growth by restricting GB migration (Zener pinning) at elevated temperatures [17,37,38]. In addition to these precipitates, MC carbides and Laves phase are also commonly observed in as-cast conditions as the result of eutectic reactions during solidification [39,40]. Much like the δ phase [13], MC carbides are shown to impede GB migration but, they can lead to intergranular cracking when present in high densities at GBs [18]. Moreover, the brittle nature of Laves phase is documented to detrimentally affect tensile strength, fracture toughness, fatigue, and elongation by serving as crack initiation sites during deformation [9,15,40,41].

Traditionally, IN718 has been manufactured using a multitude of techniques, including forging, casting, and powder metallurgical processing [42,43]. More recently, AM processes have enabled the production of complex geometries directly from digital files, providing greater design freedom and customized production [44–48]. One limitation of AM parts relative to traditionally-manufactured products is that it is impractical to strengthen AM components by work hardening [44]. Therefore, weldable, precipitation strengthened alloys like IN718 have been among the first alloys adopted to AM with notable success across multiple AM processes; e.g. powder bed fusion by laser (L-PBF) [4], powder bed fusion by electron beam (EB-PBF) [49,50], directed energy deposition by laser [51], and directed energy deposition by plasma arc [46].

The fast cooling rates ($10^3$–$10^8$ K/s) of AM processes leads to distinct microstructures in bulk AM parts compared with those produced through traditionally manufactured routes. At these cooling rates, the rejection of solute introduces constitutional supercooling ahead of the solidification front, resulting in solid-liquid interface instability and subsequent nonplanar solidification morphologies such as cellular or dendritic growth [52]. Thermo-Calc [53] simulations mimicking solidification conditions of EB-PBF processed IN718 showed segregation of Nb, Mo, Ti, and C elements to interdendritic regions and corresponding depletion from the γ matrix [54].

In addition to solidification during the deposition of a layer in AM processes, subsequent layer deposition can cause remelting and/or thermal cycling of previously solidified layers, changing the final microstructure relative to the initially deposited microstructure. For example, the layer-wise deposition of L-PBF processes can cause local tensile and compressive strain variations in as-printed IN718 that induce macroscopic residual stresses [55]. These residual stresses lead to the generation of dense dislocation forests arranged in cellular boundaries, as well as high dislocation density within the cell interiors [6,7,14]. Furthermore, the thermal cycling of AM processes has been reported to cause γ′ and γ″ precipitation in as-printed parts [56,57] Yet, there are other reports that have indicated only trace quantities of γ′ and γ″ precipitation, if any, in the as-printed condition [7,12,21]. This disparity is suggestive of variability in the as-printed microstructures based upon processing parameter and/or part geometry variations. Post processing heat treatments provide a means to control the final microstructures of AM parts despite these as-printed variations, resulting in more consistent mechanical performances. In this work, we aim to understand if heat treatment can be used to create superior AM-IN718 performance relative to traditionally manufactured IN718 materials.

## 2. Materials and Methods

### 2.1 Additively manufactured sample fabrication

Free-standing tensile specimens pursuant of ASTM E8 / E8M - 16a standard sub-size geometry were additively manufactured at 45° rotational increments with respect to the build plate normal (polar angle) and within the plane of the build plate (azimuth angle), using a Concept Laser M2 Cusing multilaser system configured with two 400W diode-pumped Yb-fiber lasers and F-Theta full quartz lenses. One of the build plates is shown in Fig. 1, with the in-plane azimuth angle convention indicated with θ. Samples were printed on P20 stainless steel build plates.

Manufacturer-provided processing parameters were used, as well as manufacturer-recommended build settings. Specifically, laser scan strategies consisting of inner skin, outer skin, and advanced contour pass methods were used to fabricate the samples (per the program provided by Concept April 2016). The inner skin consisted of a bi-direction, "serpentine" laser pass in an "island" fill pattern composed of 5 mm x 5 mm squares orthogonally oriented to each other to fill the bulk of a build layer to within 2 mm of the perimeter [58]. The outer skin was used within 2 mm of the outer perimeter of a contiguous area and was performed in a bi-directional, "serpentine" path. The advanced contour pass consisted of a single, unidirectional pass along the outer perimeter of each part to improve the surface roughness of the finished specimen. All three laser scan strategies used manufacturer-provided processing parameter, as well as manufacturer-recommended build settings: a laser power of 160 W; scanning speed of 800 mm/s; laser focus spot size 80 μm; hatch spacing 160 μm; powder layer thickness 50 μm; and 90º rotation in laser path between subsequent layers. No laser parameter modifications were made to accommodate upskin and downskin regions of the build. For all tensile specimens manufactured, the build layers through the gauge sections were fabricated using only the outer skin followed by the advanced contour pass laser scan strategies. The use of the inner skin scan strategy, in conjunction with the other two strategies, was isolated to the grip portion of the $\phi = 0º$ tensile specimens where the cross-sectional area of the build layer was sufficiently large.

The printing was performed under Ar atmosphere with the oxygen concentration maintained to 0.40 to 0.60 % of the atmosphere; the relative humidity was kept below 10%; the maximum temperature of the printing chamber did not exceed 32 °C; pre-heating of the powder bed or the build plate was not performed. Manufacturer-provided IN718 powder (gas-atomized in Ar, 10–45 µm sieved particles, marketed as CL-100NB [59]) was used entirely in a "virgin" (not recycled) condition. Chemical composition analysis of the powders and as-built samples was performed using inductively coupled plasma - atomic emission spectroscopy (ICP-AES) at NASA Glenn Research Center, and is presented in Table 1.

Table 1. Chemical composition of the IN718 powder and AM part.

|  | Ni | Cr | Fe | Nb | Mo | Ti | Al | Co | O | N |
|---|---|---|---|---|---|---|---|---|---|---|
| Powder (wt.%) | 53.13 | 19.07 | 18.23 | 5.11 | 3.06 | 0.89 | 0.42 | 0.027 | 0.020 | 0.016 |
| Build (wt.%) | 53.10 | 18.91 | 18.34 | 5.15 | 3.06 | 0.90 | 0.43 | 0.071 | 0.024 | 0.023 |

The heat treatment studies reported in this work were developed using parts printed in 45° orientations between horizontal and vertical; various θ rotations were selected. The printed tensile specimens were detached from the build plate prior to all heat treatments, with one exception detailed below. Some of these as-printed specimens were then heat treated according to AMS 5662 [11], in which a solution anneal at 980 °C for 1h was followed by air cooling to room temperature and a 2-step aging treatment: 720 °C for 8h followed by furnace cooling down to 620 °C within 2 h, where the temperature is held for 8 h before air cooling to room temperature (standard heat treatment, SHT). Residual stress introduced by the AM process and thermal contact with the build plate may impact microstructure evolution during heat treatment. To study this effect, the standard heat treatment was applied before (SHT-1) and after (SHT-2) the samples are removed from the build plate.

Other heat treatments were then designed to isolate specific aspects of the thermodynamics and kinetics driving microstructural evolution that were not completely understood from the SHT studies alone (as further discussed in Sections 3 and 4). Specifically, some of the as-printed samples (first removed from the build plate) were "direct-aged" at either 620 ºC for 24 h or 720 ºC for 24 h; that is, with no solution anneal. Finally, a modified two-step anneal-then-age heat treatment was designed using knowledge of the SHT and direct-age heat treatments and their effects on microstructures and properties. Specifically, anneal-then-age samples were water quenched after a 15 min solution anneal at 1020 ºC, then aged at 720 ºC for 24 h.

A summary of the heat treatments studied in this work is provided in Table 2, together with the acronym designation used to refer to the heat treatments throughout this article. For all samples used to compare the different heat treatments, the as-printed surfaces of the samples were not machined, polished, or otherwise modified. To examine the effect of surface finish on the best reported properties, one sample that underwent the SA1020+A720 treatment was machined prior to mechanical testing, and is referred to as SA1020+A720+M. Each surface of this E8 sub-size tensile specimen (with gauge section dimensions of 5.96 mm × 2.10 mm) was machined using an end mill fitted with a 5 mm carbide bit followed by light grinding with 1000 grit SiC paper for the finished surface. The geometric ratios from the original tensile specimen were maintained, with the final gauge section dimension of 5.15 mm × 1.38 mm.

Table 2. Heat treatment schedules for various conditions examined in this work. "FC" stands for furnace cooling and "AC" for air cooling.

| Condition | Designation | Solution Anneal | Aging |
|---|---|---|---|
| As-printed | AP | -- | -- |
| Solution annealed | SA980 | 980 ºC/1 h/WQ | -- |
| Standard heat treatment | SHT-1* | 980 ºC/1 h/AC | 720 ºC/8 h/FC 50 ºC/h+620 ºC/8 h/AC |
| Standard heat treatment | SHT-2 | 980 ºC/1 h/AC | 720 ºC/8 h/FC 50 ºC/h+620 ºC/8 h/AC |
| Direct age: 620 ºC, 24 h | DA620 | -- | 620 ºC/24 h/AC |
| Direct age: 720 ºC, 24 h | DA720 | -- | 720 ºC/24 h/AC |
| Solution annealed + aged | SA1020+A720 | 1020 ºC/0.25 h/WQ | 720 ºC/24 h/AC |

*Tensile specimen connected to build plate during heat treatment

## 2.2 Wrought sample fabrication

To compare the AM vs. wrought materials properties, an ASTM E8 tensile sample with a gage cross sectional area of 31.67 mm$^2$ was machined from a wrought plate and heat treated using the SHT schedule described in Section 2.1 (

Table *2*) by Metals Technology, Inc.

## 2.3 Mechanical testing

Monotonic tensile testing of all examined conditions of the AM samples was performed using an MTS servo-hydraulic load frame equipped with an MTS 662.20H-05 load cell. A cross-head speed of 4.5 mm/min was used, corresponding to an approximate strain rate of 0.003 s$^{-1}$. Strains were measured using an extensometer (MTS model 634.12F-24). All engineering stress-stress values for the AM and wrought samples were converted to logarithmic (i.e., "true") stress and strain [60].

The wrought sample was also tested following the ASTM E8 standard. The test data was provided by Metals Technology, Inc.

**2.4 Microstructure characterization**

Samples for optical microscopy (OM) and electron backscatter diffraction (EBSD) were prepared using standard metallographic polishing techniques with a final polishing step with 0.02 colloidal silica on a VibroMet vibratory polisher for 24 h. OM samples were etched with glyceregia (3 parts HCl, 2 parts glycerol, 1 part $HNO_3$) then imaged using a Keyence VHX-5000 digital microscope. ESBD measurements were performed using a FEI Helios Nanolab 600i DualBeam SEM/FIB configured with an EDAX Hikari Super EBSD camera. The statistics of grain size widths and lengths reported in this work consider analysis of at least 100 grains per condition from OM and/or EBSD data. All EBSD data was processed using EDAX OIM Analysis 7 software (version 7.2.1). The Taylor factor (M) was calculated using $\{1\bar{1}1\}\langle 110\rangle$ fcc slip systems and the deformation gradient tensor, $\boldsymbol{F} = \begin{bmatrix} -0.5 & 0 & 0 \\ 0 & 1 & 0 \\ 0 & 0 & -0.5 \end{bmatrix}$.

Conventional bright-field (BF), dark-field (DF), high-angle annular dark-field (HAADF) and high-resolution (scanning) transmission electron microscopy (HRTEM and HR HAADF STEM) characterizations were performed using a FEI Talos TEM (FEG, 200 kV equipped with ChemiSTEM X-ray energy dispersive spectroscopy (EDX) four detector technology) and a FEI Titan 80-300 with electron-probe Cs-correction operated at 300 kV. Additionally, an FEI Titan G2 80–200 with ChemiSTEM technology was also used for high-spatial-resolution EDX mapping. TEM samples were prepared from the gauge sections of the 45° tensile bars by equally grinding both surfaces down to a final thickness of 90–100 μm; a mechanical punch was then used to extract discs with a diameter of 3 mm. A Fischione automatic twin-jet electropolisher (Model 120) at 15

V and an electrolyte of 10 vol% perchloric acid in methanol at −32 °C was then used to further thin the TEM foils. Lattice dislocation densities of cell interiors were measured in BF TEM micrographs by line intercept method [61] using at least 50 dislocation cells. Entangled forests of dislocations along the cell boundaries were excluded from this calculation – dislocation cell sizes were separately analyzed and reported. Average precipitate sizes, interparticle distances, and corresponding standard deviations are reported using measurements of approximately 50 precipitates from several HRTEM, HR HAADF STEM, BF and DF images taken from various regions of each sample.

## 3. Results

The presentation of results is provided in two subsections. The mechanical properties are presented in Section 3.1. Microstructures of the AM materials are presented in Section 3.2. Process-structure-property correlations are discussed in Section 4.

### 3.1. Monotonic mechanical properties

The anisotropy and statistical variability in the mechanical properties of as-printed (AP) samples (Fig. 1) due to both build orientation (a-d) and azimuthal angle (e-h) are illustrated in Fig. 2. Relevant to this report, note that samples with "average" AP mechanical properties were chosen to understand the SHT and to develop new heat treatments.

Fig. 3 shows the monotonic tensile properties of AM-IN718 in the AP, SA980, SHT-1, SHT-2, DA620, DA720, and SA1020+A720 heat treatment conditions (refer to Table 2 for acronym designations), as well as the wrought IN718 in the SHT condition. Fractures of all samples occurred within the gauge length of extensometer. Table 3 lists the yield strength, UTS, elongations-to-failure extracted from the true stress-strain curves (Fig. 3), and the relative differences of the AM samples compared to the cast and wrought sample (SHT condition), as well as compared to the AP sample. These results show that the yield strength of the AM material decreases by ~18% after the solid solution anneal (SA980), but is otherwise improved from the AP condition using any of the investigated heat treatments according to their full schedules. Specifically, relative to the AP condition, the yield strength increased 47% using DA620, 49% using SHT-1, 63% using SHT-2, 71% using DA720, and 64% using SA1020+A720 heat treatments. In contrast, the variation of UTS values shows that the AP and SA980 conditions have similar UTS despite an ~180 MPa difference in yield strength, indicating a higher strain hardening for the SA980 sample. More interestingly, while the DA720 and SA1020+A720 heat treatments achieve a higher yield strength than the SHT-1 (15% and 10%, respectively) and wrought SHT samples (12% and 7%, respectively), the four materials have comparable UTS due to the higher

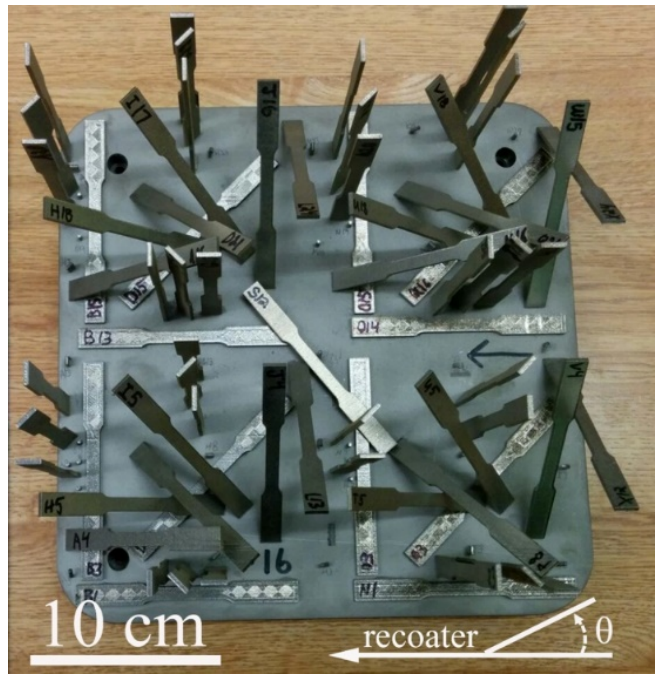

Fig. 1. Build plate of free-standing tensile specimens demonstrating various build orientations and azimuthal angles (θ) used in their construction.

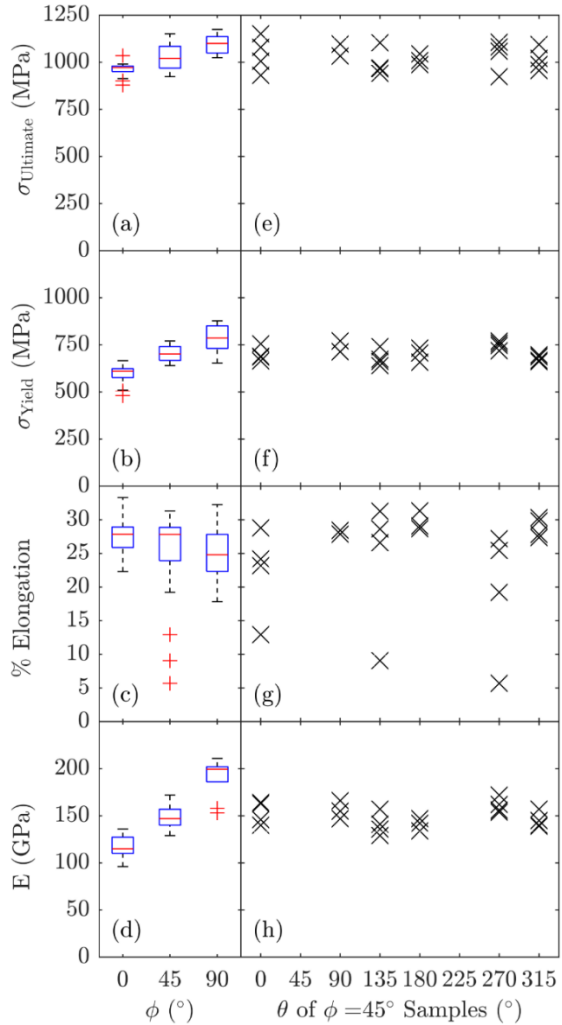

Fig. 2. Statistical variability of ultimate tensile strength σUltimate, yield strength σYield, % elongation to failure, and Young's modulus (E) for the free-standing tensile specimens with respect to build orientation angle ϕ, where 0° is vertical and 90° is flat on the plate is given in (a-d) as box plots, where the red lines within the box represent the median values of the normal distributions of n = 33 tests for the 0° samples, n = 22 for 45° and n = 10 for 90°. The boxes show the extent of the 2nd and 3rd quartiles of those same distributions, while the error bars show

the extent of the 1st and 4th quartiles and outliers are plotted as red crosses. Then, variability of the ϕ = 45° samples used in this study as a function of rotation with respect to the recoater blade θ as defined in Fig. 1 is shown in (e-h), where individual data points are plotted.

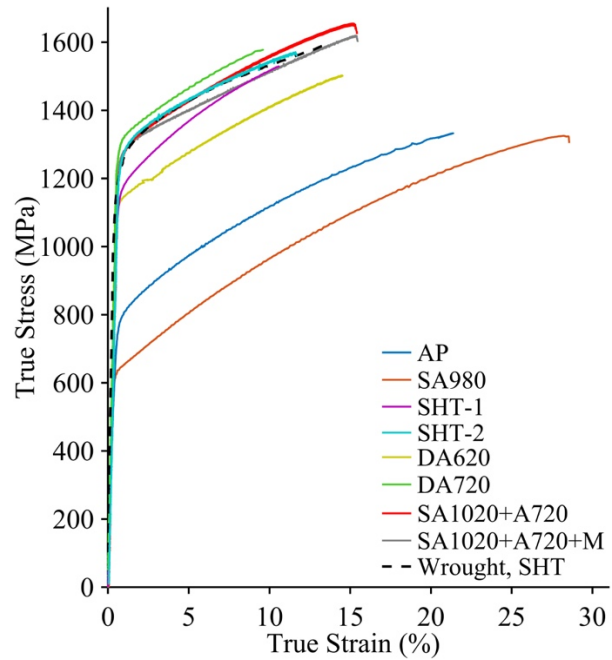

Fig. 3. Monotonic tensile behaviors of IN718 produced by L-PBF then heat treated as indicated in Table 2, in addition to a machined sample of the SA120+A720 condition (indicated SA120+A720+M) and wrought IN718 in the SHT condition (Wrought, SHT).

Table 3. Yield strengths ($\sigma_{YS}$), ultimate tensile strengths ($\sigma_{UTS}$) and elongations to failure of AM-IN718 subjected to the various heat treatments studied in this paper (see Table 2). In addition to the values, the relative percent (%) differences relative to wrought, standard heat treated ($\Delta^{W,SHT}$) and as-printed ($\Delta^{AP}$) IN718 are also given.

| Condition | $\sigma_{YS}$ | | | $\sigma_{UTS}$ | | | Elongation | | |
|---|---|---|---|---|---|---|---|---|---|
| | (MPa) | $\Delta^{W,SHT}$ | $\Delta^{AP}$ | (MPa) | $\Delta^{W,SHT}$ | $\Delta^{AP}$ | (%) | $\Delta^{W,SHT}$ | $\Delta^{AP}$ |
| AP | 760 | -34.5 | - | 1335 | -17.1 | - | 21.3 | +57.8 | - |
| SA980 | 620 | -46.6 | -18.4 | 1325 | -17.7 | -0.75 | 28.6 | +112 | +34.3 |
| SHT-1 | 1135 | -2.16 | +49.3 | 1530 | -4.97 | +14.6 | 10.6 | -21.5 | -50.2 |
| SHT-2 | 1240 | +6.90 | +63.2 | 1560 | -3.11 | +16.9 | 11.6 | -14.1 | -45.5 |
| DA620 | 1120 | -3.45 | +47.4 | 1500 | -6.83 | +12.4 | 14.5 | +7.41 | -31.9 |
| DA720 | 1300 | +12.1 | +71.0 | 1580 | -1.86 | +18.4 | 9.6 | -28.9 | -54.9 |
| SA1020+A720 | 1245 | +7.33 | +63.8 | 1640 | +1.86 | +22.8 | 16.6 | +23.0 | -22.1 |
| Wrought, SHT | 1160 | - | +52.6 | 1610 | - | +20.6 | 13.5 | - | -36.6 |

amount of strain hardening exhibited by the SHT-1 and wrought SHT samples. The SA1020+A720 sample exhibits the highest UTS (1640 MPa) of all the aged conditions evaluated.

Compared to the AP condition, solution annealing by itself (SA980) improved elongation to failure, but aging reduces elongation to failure by 22-55%. The largest decrease in elongation (55%) was observed in the DA720 condition, while the SA1020+A720 condition yielded the best elongation of all the aged conditions examined (16.6%). In contrast, the SHT-1 and DA620 AM samples have comparable YS and UTS, but the greater elongation of the DA620 sample results in a lower rate of strain hardening (with respect to the total accumulation of strain). Moreover, comparing the two SHT conditions, despite having higher YS in the SHT-2 condition, the elongation was still 9% higher over the SHT-1 condition, indicating an improvement over usual strength vs. elongation trade-offs that is further discussed in Section 4.

## 3.2. Microstructure characterizations

Table 4 summarizes the sizes of the secondary phases, dislocation cells and dislocation densities within the cells of the AM-IN718 materials (again, see Table 2 for definition of the different conditions). The following subsections present the microstructures specific to each processing condition, including detailed quantification of the data presented in Table 4.

### 3.2.1. The as-printed (AP) microstructure

Dark, sweeping lines in the optical micrograph shown in Fig. 4(a) are the cross-sections of weld pools generated during deposition of each layer. Columnar grain growth that spans these weld pool boundaries is evident in Fig. 4(a) to Fig. 4(b). This result and the absence of lack-of-fusion defects provides evidence that in subsequent build layers the laser fully penetrates the depth of the previous layer melt pool, and that the unmelted grains at the weld-pool/deposition interface serve as nucleation sites for competitive epitaxial growth, as has been previously established for L-PBF [62]. Fig. 4(b) further shows that these columnar grains have misorientation with respect to each other due to different growth directions, considering the orientation map colored according to the inverse pole figure (IPF) made with respect to the plane normal parallel to the tensile axis (TA). Low-angle grain boundaries (LAGBs) and high-angle grain boundaries (HAGBs), as defined by thresholds of $3º \leq \theta \leq 15º$ and $\theta \geq 15°$ misorientations, respectively, are highlighted by red and black lines, respectively. The columnar grains are shown to be aligned with the build direction. The average grain dimensions were measured to be $13.4 \pm 9.4$ μm in width and $96.1 \pm 72.9$ μm in length when including both LAGBs and HAGBs. Moreover, when only considering HAGBs, the average grain dimensions were $25.2 \pm 19.7$ μm in width and $108.4 \pm 64.3$ μm in length.

Table 4. Statistical measurements of AM-IN718 after application of various heat treatments.

|  | AP | SA980 | SHT-1 | SHT-2 | DA620 | DA720 | SA1020+A720 |
|---|---|---|---|---|---|---|---|
| **Dislocation cell statistics** | | | | | | | |
| Dislocation cell size (nm) | 620 ± 180 | 650 ± 125 | ** | 610 ± 140 | 640 ± 110 | 620 ± 110 | 660 ± 160 |
| Dislocation density ($10^{14}$ m$^{-2}$) | 1.6 ± 0.8 | 0.86 ± 0.24 | ** | ** | 1.4 ± 0.4 | 0.24 ± 0.06 | ** |
| **Precipitate size (nm)** | | | | | | | |
| Laves phase | 214 ± 62 | ** | ** | ** | 234 ± 46 | 240 ± 58 | ** |
| δ phase – Major axis | ** | 500 ± 140 | 940 ± 250 | 700 ± 290 | ** | ** | ** |
| δ phase – Minor axis | ** | 83 ± 42 | 110 ± 35 | 134 ± 27 | ** | ** | ** |
| γ″ phase – Major axis | * | ** | 64 ± 28 | 27 ± 6 | 8 ± 2 | 31 ± 8 | 29 ± 7 |
| γ″ phase – Minor axis | * | ** | 23 ± 7 | 8 ± 2 | 3 ± 1 | 10 ± 3 | 8 ± 2 |
| γ′ phase | * | ** | 25 ± 6 | 18 ± 3 | ** | 21 ± 7 | 23 ± 4 |
| γ′/γ″ co-precipitates | ** | ** | 24 ± 6 | 16 ± 4 | ** | 15 ± 2 | 18 ± 3 |
| **Interparticle spacing (nm)** | | | | | | | |
| Laves phase | 316 ± 103 | ** | ** | ** | 399 ± 96 | 380 ± 105 | ** |
| δ phase | ** | 575 ± 305 | 700 ± 375 | 537 ± 384 | ** | ** | ** |
| Strengthening precipitates | ** | ** | 24 ± 11 | 15 ± 6 | 7 ± 2 | 15 ± 4 | 16 ± 5 |

*Indistinguishable due to small size **Feature not observed

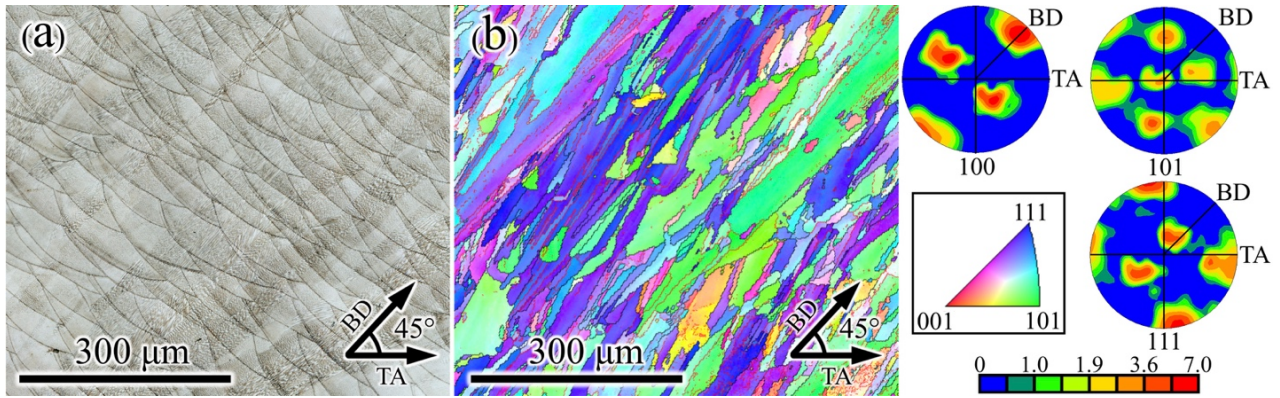

Fig. 4. (a) Optical micrograph of the side view of a 45º-oriented tensile specimen is given for the as-printed (AP) condition. Melt pool boundaries are indicated by the darker, swooping lines that are normal to the build direction. (b) The crystallographic orientation map is presented colored according to the inverse pole figure (IPF) that is given with respect to the plane normal (($hkl$)s) parallel to tensile axis (TA). The build direction (BD) and tensile axis of the sample lie in the plane of the page, 45º apart, as indicated by the drawn coordinate systems. LAGBs and HAGBs are outlined with red and black lines, respectively. {100}, {101}, and {111} pole figures with intensities colored by multiples of random distribution according to the indicated scale.

Moreover, the grain morphology is observed to vary even though the melt pool morphology remains more-or-less uniform – smaller, more equiaxed grains are interspersed and sometimes cluster in between or within the columnar grains. The inset {100} pole figure (PF) indicates a strong fiber of the texture with {100} aligned with the build direction (BD), which lies 45º with respect to the tensile axis (TA) of the specimen. Two additional fibers components have {100} oriented in transverse-to-build orientations, misaligned from the major axes of the 2 mm x 6 mm rectangular cross-section. The 3-fiber-component nature to the texture is confirmed considering the {110} and {111} pole figures, noting that one of the fibers has {111} aligned with the TA, and another {101}, consistent with the orientation map in Fig. 1(b). This crystallographic texture is consistent with the preferential growth direction along ⟨001⟩ for fcc metals [63] and previous reports of AM-IN718 [15,58,64].

Color gradients observed within individual grains (Fig. 4(b)) indicate low-angle misorientations. The origin of these misorientations is revealed considering the conventional BF TEM micrographs of Figs. 5, 6(a) and 6(b). These figures show that each grain contains a columnar, cellular substructure. The cross-sections of the cells are $620 \pm 180$ nm across. Lattice dislocation densities within the cell interiors were measured to be $1.6 \pm 0.8 \times 10^{14}$ m$^{-2}$ (Table 4). Additionally, lattice fringes corresponding to nanoprecipitation within the cell interiors were observed, as indicated with arrows in Fig. 6(c), and measured to be $4 \pm 1$ nm. However, due to their very small size (overlapping with the matrix) and sparsity, fast Fourier transform (FFT) and selected area electron diffraction (SAED) techniques were not able to confirm their phases.

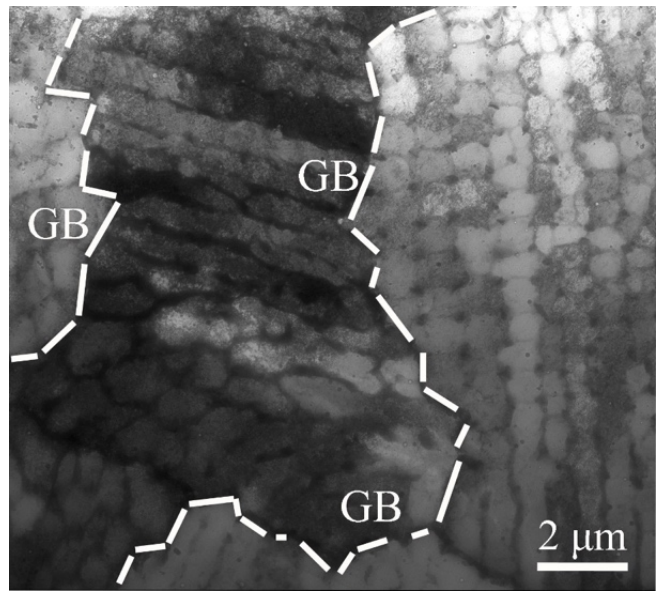

Fig. 5. Conventional BF micrograph shows the morphology of grains and the existence of submicron cellular substructure within each grain.

Chemical mapping using STEM-EDX was performed on the region of the sample shown in the HAADF-STEM micrograph in Fig. 6(d). The uniformly brighter regions at intercellular boundaries (Fig. 6(b)) coincide with Nb and Ti enrichment (Fig. 6(e-f)). In contrast, the heterogeneous bright spots (Fig. 6(d)) indicate the existence of particles at the intercellular boundaries. These particles are consistent with three known phases in IN718: metal carbides, metal oxides, and Laves. Specifically, MC carbides are known to be rich in Nb and Ti and depleted in Ni [65], which is consistent with the particles evident to be Nb and Ti-enriched and Ni-depleted Fig. 6(e-g). The Al- and O-rich particles shown in Fig. 6(h-i) confirm the presence of oxides, which have been reported in laser-based AM processes of various alloys including IN718 due to O entrapment [46,66].

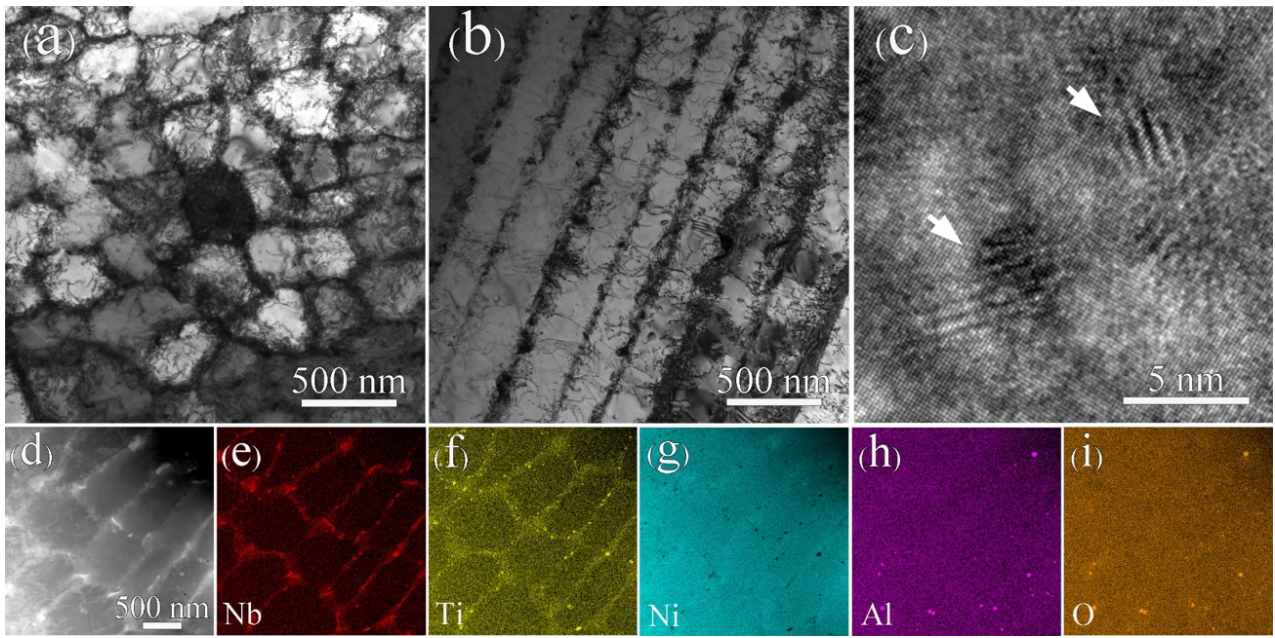

Fig. 6. (a,b) Conventional BF micrographs of the as-printed IN718 showing formation of columnar dislocation cells. (c) High-resolution TEM micrograph of nanoprecipitates observed within the dislocation cells. (d) HAADF-STEM micrograph and (e-i) corresponding STEM-EDX maps highlighting segregated elements at cell boundaries.

The irregularly shaped particles of Nb-rich phases (Fig. 6(d-e)) are consistent with the reported composition of Laves phase [65,67]. The magnified BF and corresponding SAED shown in Fig. 7(a) provides further evidence of the Laves phase. The superlattice reflections in SAED pattern in the inset of Fig. 7(a), which was taken along the $[111]_\gamma$ // $[0001]_{Laves}$ zone axis, belong to the hexagonal close-packed (hcp) structure of the Laves phase. Fig. 7(b) shows the corresponding central dark-field micrograph of Fig. 7(a) that was imaged using the **g** = $(\bar{1}010)_{Laves}$ reflection indicated with the white circle in the inset of Fig. 7(a). The irregularly shaped particles along the cell boundaries highlighted in Fig. 7(b) share similar morphology to the Nb-rich particles in Fig. 6(e). These Laves particles measure to be 214 ± 62 nm, with an interparticle distance of 316 ± 103 nm.

This as-printed microstructure is consistent with previous reports of L-PBF AM-IN718 [4,12–15,17,20]. Chemical segregation due to elemental solutes being ejected to dendrite surfaces encourages localization of oxide, carbide, and Laves particle formation upon the cellular dislocation substructures as a result of rapid cooling rates during additive manufacture, as recently discussed by Yoo et al. [20]. These observations give direct evidence for the as-printed Ti and Nb segregation they hypothesized in observing solution-annealed AM-IN718 microstructures, and is consistent with previous reporting [17,21,57,68].

*3.2.2. The solution annealed (SA980) microstructure*

Fig. 8(a) shows a BF micrograph of the AM-IN718 after a solution annealing at 980 °C for 1 h. It is obvious that large, needle-like, Nb-rich δ phase (indicated by white arrows), with an average length and width of 500 ± 140 nm and 50 ± 20 nm, respectively, have formed at the intercellular boundaries and span through the cell interiors. Their formation is expected, as the solution anneal temperature is within the δ phase nucleation temperature range of 700 °C to 1000 °C [27,28]. In addition to the appearance of these δ precipitates that were largely absent in the AP condition, the dislocation density within cell interiors (as in Fig. 8(b)) is diminished, and is measured to be $0.86 \pm 0.24 \times 10^{14}$ m$^{-2}$, approximately half the dislocation density observed of the AP condition. However, the walls of the cellular structures, while less entangled and populated with apparently fewer dislocations, have not annihilated; the cells themselves still measure to be $650 \pm 125$ nm – statistically unchanged in size relative to the AP condition. Fig. 8(b) and (c) present a HAADF-STEM micrograph with corresponding Nb map of AM- IN718 after solution anneal.

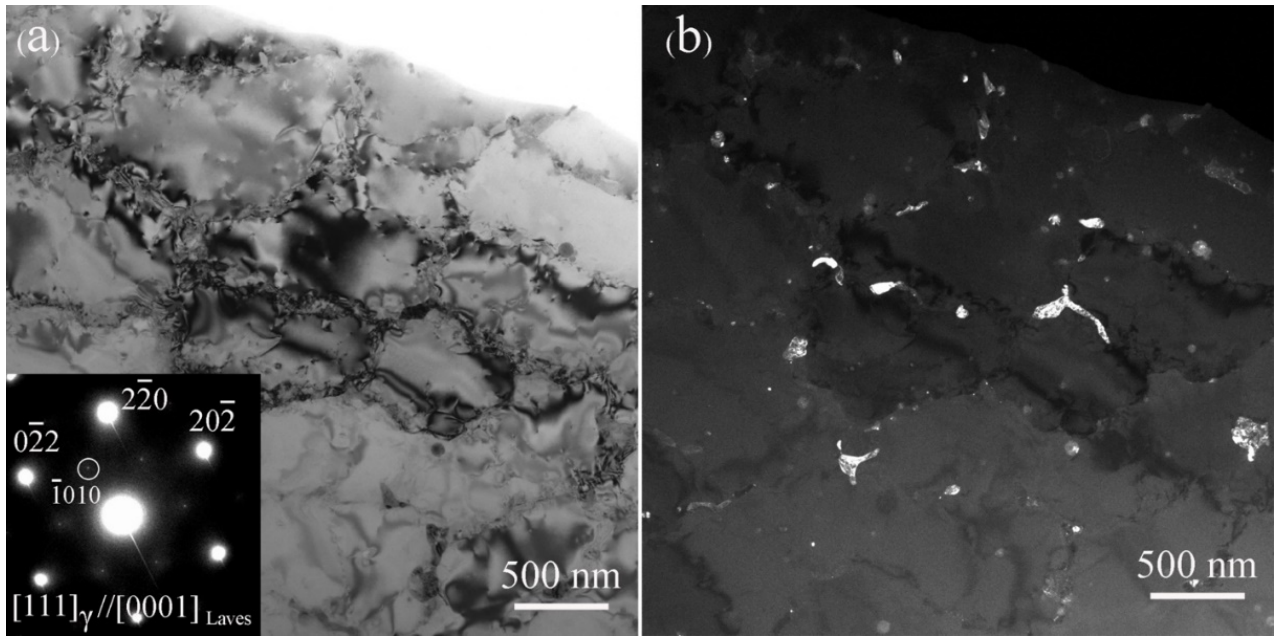

Fig. 7. (a) Conventional BF micrograph of as-printed IN718, with the corresponding SAED pattern along the $[111]_\gamma$ // $[0001]_{Laves}$ zone axis shown as an inset. (b) Conventional central DF micrograph showing the Laves phases using $\mathbf{g} = (\bar{1}010)_{Laves}$ reflection indicated by the white circle in the SAED pattern given in (a).

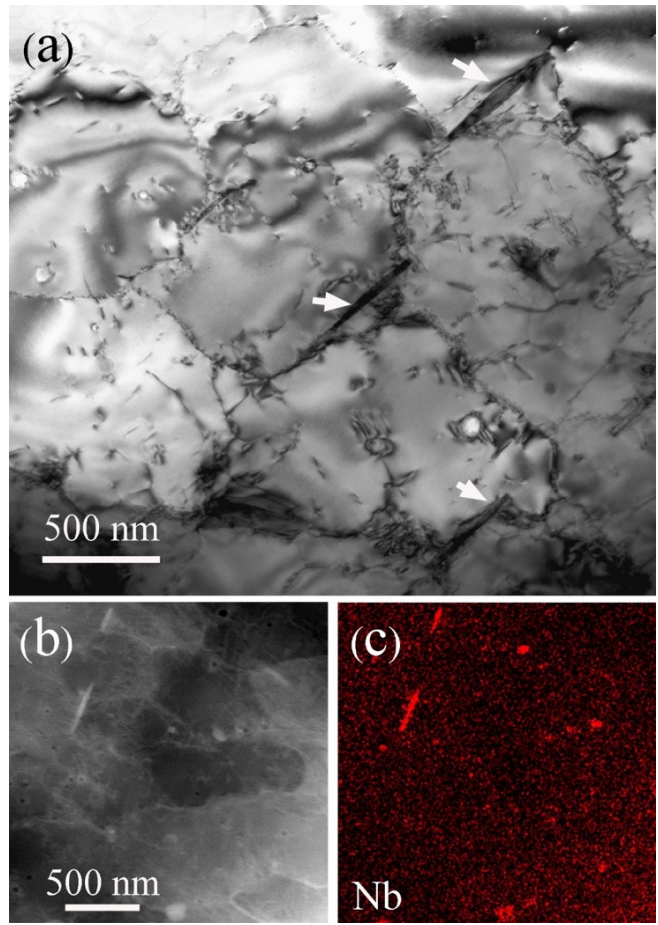

Fig. 8. (a) Conventional BF micrograph of L-PBF IN718 after SA980 heat treatment. The white arrows indicate the formation of δ phase. (b) HAADF-STEM and (c) corresponding Nb mapping using the STEM-EDX technique showing homogenization of Nb throughout the microstructure.

The absence of uniformly bright regions around cell boundaries in Fig. 8(b-c) indicates that the solution annealing leads to homogenization of the previously segregated Nb and Ti back into the matrix. Moreover, Laves phase are not observed, indicating that the solution annealing temperature and time (i.e. 980 °C for 1 h) was sufficient to dissolve the eutectic Laves phase. However, the carbides and oxides did not dissolve. Finally, the nanoprecipitates suggested by fringes observed of HR-TEM images of the AP condition (Fig. 6(c)) are not observed in samples of this condition (not shown), indicating that they were dissolved into the matrix through the solution anneal, as expected. Indeed, this microstructure is similar to that previously reported for solution annealed AM-IN718 at 1065 °C [20], only here, δ precipitates are also observed.

*3.2.3. The standard heat treatment applied to entire build (SHT-1) microstructure*

Fig. 9 (a) and (b) show low magnification HAADF-STEM and magnified BF micrographs of AM IN718 after SHT-1. Recall that during the SHT-1 heat treatment the specimen remains attached to the build plate. Relative to SHT-2, the SHT-1 heat treatment results in slower cooling rates in between each heat treatment step due to the extra thermal mass of the build plate itself, as well as other samples, during heat treatment. Needle-like δ precipitates can be seen at both intercellular and grain boundaries; they have an average length and width of 940 ± 250 nm and 110 ± 35 nm, respectively. More interestingly, the entangled dislocations inherent to the cell walls of the AP condition have recovered. Dislocation recovery was most likely enhanced by the nucleation and growth of the coarse $\gamma'$ and $\gamma''$ precipitates that are configured in cellular arrangements of the same size (630 ± 75 nm), seemingly in place of the dislocation entanglements (Fig. 9(a) and (b)). Dislocations are known to serve as $\gamma'$ and $\gamma''$ nucleation sites [69,70]. Quantification of the nanoprecipitates within this microstructure shows that three scales have formed: 1) the 160 ± 37 nm long lenticular precipitates that mark the cell boundaries (Fig. 9(a)), while 2) 68 ± 28 nm long lenticular and 3) 25 ± 6 nm equiaxed precipitates have formed within the cells (Figure 10). This disparity in precipitate sizes between the cell walls and interiors suggests that the dislocations along the cell boundaries after the solution anneal can promote 1) the heterogeneous nucleation of precipitates on boundaries [71] due to high local strain and 2) higher growth rate of precipitates by increased elemental mobility via pipe diffusion [72]. The corresponding SAED pattern in the inset of Fig. 9 (b) taken along the $[001]_\gamma$ zone axis exhibits the superlattice reflections consistent with $\gamma'$ and $\gamma''$ precipitates that have $(001)_{\gamma''} // \{001\}_\gamma$ and $[001]_{\gamma'} // <001>_\gamma$ orientation relationships and similar lattice constants [24,26,28,73]. As such, all superlattice reflections of $\gamma'$ overlap with those from $\gamma''$ (indicated with solid circles in the SAED pattern). The other superlattice reflections of $\gamma''$ are identifiable and indicated by dashed circles in the SAED pattern. Therefore, the presence of $\gamma''$ is confirmed, while the presence of $\gamma'$ precipitates is ambiguous considering only the SAED pattern.

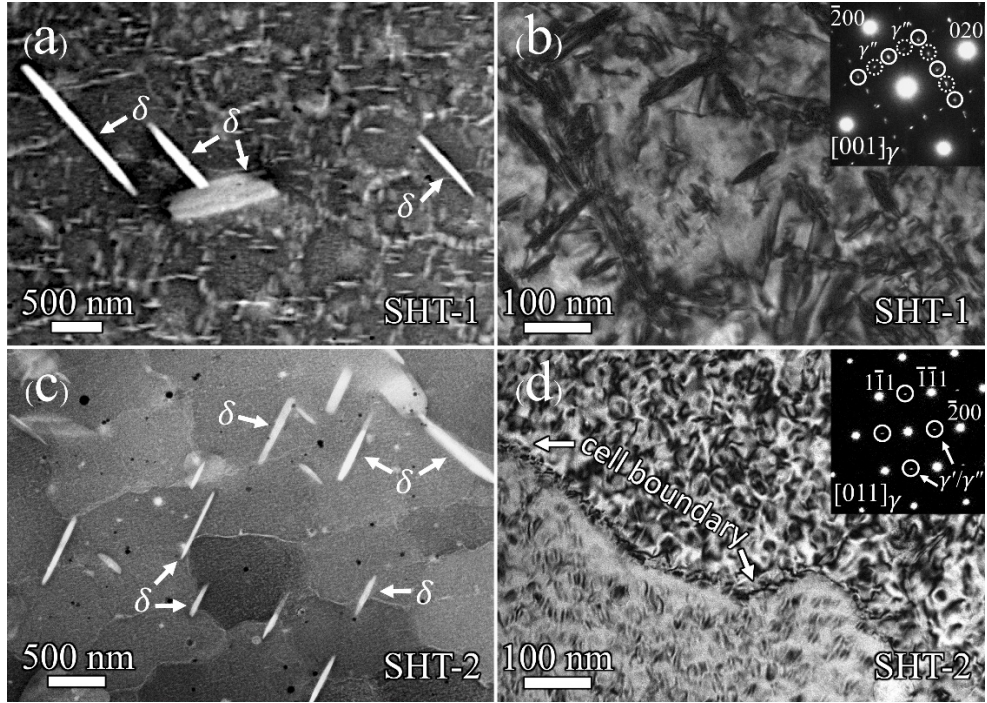

Fig. 9. (a) HAADF-STEM micrograph showing cell-like substructure defined by lenticular precipitates and δ phase precipitates located along intercellular boundaries (indicated by white arrows) after SHT of tensile specimens attached to the build plate (SHT-1). (b) Magnified BF micrograph of cell interior and boundary. Corresponding inset SAED pattern of γ′, γ″ precipitates along the $[001]_\gamma$ zone axis. The solid circles correspond to superlattice reflections that belong to both γ′ and γ″, while spots outlined by the dashed circles originate only from γ″. (c) Conventional BF micrograph after SHT of detached tensile specimen (SHT-2) showing retained dislocation cell substructures, no observed lattice dislocations, and δ phase precipitates located along intercellular boundaries (indicated by white arrows). (d) Magnified BF micrograph of dislocation cell boundary free of large, heterogeneously nucleated γ″ precipitates. Dense nanoprecipitation is observed within the cell interiors by diffraction contrast. Corresponding inset SAED pattern of γ′, γ″ precipitates along the $[011]_\gamma$ zone axis. The solid circles correspond to superlattice reflections that belong to both γ′ and γ″ phases.

Thus, high–spatial resolution STEM-EDX mapping was performed to distinguish the presence and morphology of $\gamma'$ precipitates from $\gamma''$ in the SHT-1 condition. Fig. 10(a) shows a HAADF-STEM micrograph, and corresponding elemental maps are presented in Fig. 10(b–d). The larger Nb-rich, lenticular particles (Fig. 10(b)) and the Al-rich spherical particles (Fig. 10(c)) are consistent with previous reports of the composition and morphology for $\gamma''$ and $\gamma'$, respectively [74]. Confirming the presence of both $\gamma'$ and $\gamma''$ after the two-step aging treatment.

Moreover, coprecipitates are indicated by white arrows in the magnified HAADF-STEM micrograph of Fig. 11(a) and measure at $24 \pm 6$ nm in length. In examining the HR HAADF STEM image of Fig. 11(b), the $\gamma''$ lattice structure is distinguishable from the $\gamma$, but a separate $\gamma'$ lattice structure is not obvious. However, note that the top side of the $\gamma'$ and $\gamma''$ interface is rounded and not atomically planar, while the bottom interface of the $\gamma''$ phase is atomically planar. It is known that one of the three coherent, coprecipitate variants of the $\gamma'$ and $\gamma''$ phases has an atomically planar

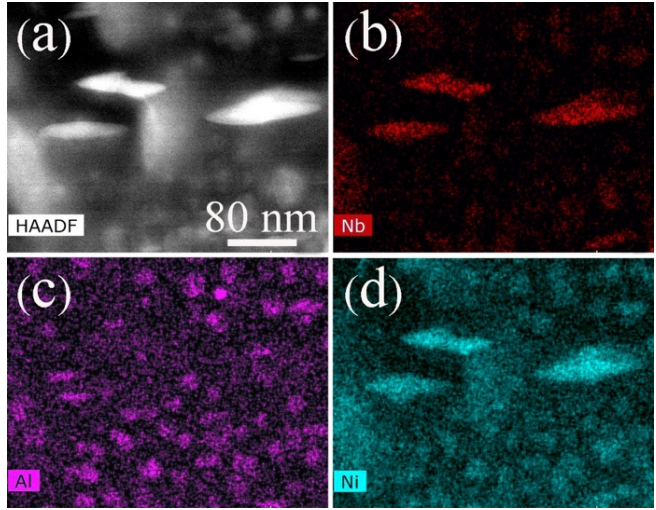

Fig. 10. (a) HAADF-STEM micrograph showing presence of nanoprecipitates in AM-IN718 after SHT-1. (b–d) Corresponding elemental STEM-EDX map to differentiate between Al-rich γ′ and Nb-rich γ″ precipitates.

$[001]_{γ″}$ interface that can be viewed edge-on in this $<110>_γ$ viewing direction [23]. Thus, the observation of this atomically-flat interface definitively indicates their presence. Furthermore, there is an atomic layer rich in Al near the interface in γ′/γ″ coprecipitates [75], although this feature was not directly verified in the present study. The ends of this layer are marked by arrowheads, and the layer is definitively dark relative to the other atomic layers in HR HAADF STEM micrograph of Fig. 11(b), consistent with an Al-rich layer. (Because the intensity of the atoms in STEM mode is proportional to their atomic number, the Al single layer appears as a low-intensity layer compared to its neighbors.) Al is known to be unfavorable in γ″ structure because of the size discrepancy between Nb and Al atoms and the fact that they occupy the same ordered lattice sites in γ″ precipitates. Still, it is difficult with HR HAADF STEM alone to distinguish the

γ′ portion of the coprecipitate from the surrounding γ matrix since the superlattice ordering is relatively weak, and the net atomic number for γ and γ′ phases of IN718 is similar [23].

Therefore, high–spatial resolution STEM-EDX mapping was also performed on a single-sided coprecipitate, and the combined Al and Nb elemental map is presented in Fig. 11(c). Using this map, the Al-rich γ′ portion of the coprecipitate can be easily distinguished from both the matrix and the Nb-rich γ″ side. Hence, we conclude that in addition to the presence of monolithic γ′ and γ″ precipitates (Fig. 9 and Fig. 10), coprecipitates are also present in the SHT-1 condition of L-PBF AM-IN718.

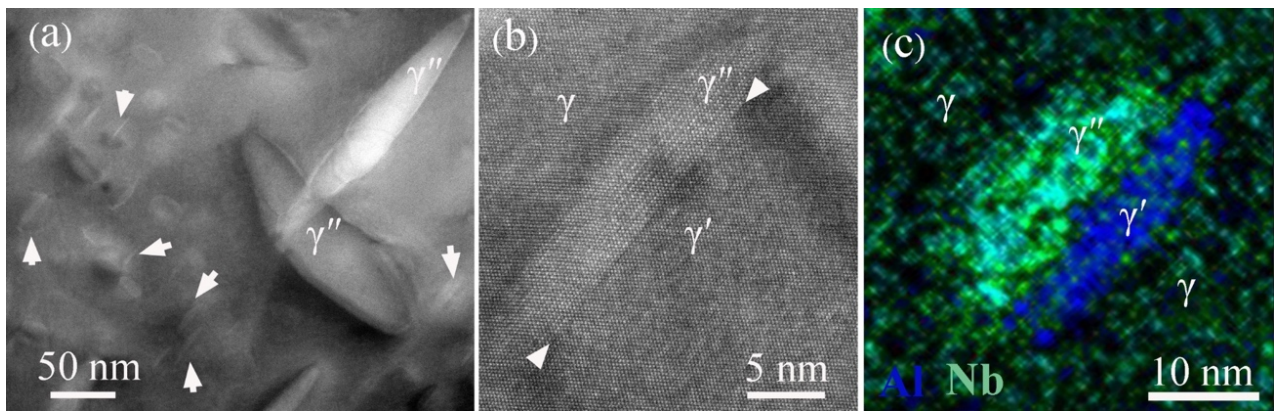

Fig. 11. (a) High-magnification HAADF-STEM micrograph of AM-IN718 after SHT-1 showing the existence of coprecipitates (indicated by arrows). The zone axis is <110>$_γ$. (b) HR HAADF STEM micrograph of single-sided coprecipitate exhibiting a planar {001} interface that is characteristic of the γ′/γ″ interface of the coprecipitates. Arrowheads indicate an Al-rich layer between γ′ and γ″ portions. (c) Corresponding STEM-EDX map of the coprecipitate showing combined Nb and Al elemental map.

*3.2.4. The standard heat treatment applied to individual samples (SHT-2) microstructure*

Fig. 9 (c) and (d) show low magnification HAADF-STEM and high magnification BF micrographs, respectively, of AM IN718 after applying SHT-2 conditions. Compared to samples heat treated while still attached to the build plate, SHT-2 samples are expected to experience higher cooling rates after each heat treatment step. Quantitative analysis (Table 4) shows a slight decrease in δ precipitate size compared to SHT-1 (average length and width of 700 ± 290 nm and 134 ± 27 nm, respectively). More profoundly, heterogeneous nucleation of γ′ and γ″ precipitates along cell boundaries is not detected; instead, subgrain boundaries remain, well-defined by dislocations and cellular structure of the same size as the AP condition (610 ± 140 nm). This marked difference between SHT-1 and SHT-2 microstructures results from the differences in cooling rates in between heat treatment steps strongly influencing the precipitation kinetics, as was established in studying traditionally manufactured nickel superalloys [70].

The corresponding SAED pattern in the inset of Fig. 9 (d) taken along the $[011]_\gamma$ zone axis again exhibits the superlattice reflections consistent with γ′ and γ″. Similar analyses for SHT-1 confirm the phases and morphologies of nanoprecipitates given in Table 4 for the SHT-2 condition. The primary differences between the cell interiors for SHT-2 vs. SHT-1 are that the nanoprecipitates are not as coarse (SHT-2 statistics: γ″: 27 ± 6 nm; γ′: 18 ± 3 nm; γ″/γ′ coprecipitates: 16 ± 4 nm), especially for the γ″ phase, which is 1/3 the size relative to the SHT-1 condition. Concurrently, the nanoprecipitates are denser at 15± 6 nm spacing.

*3.2.5. The direct aged at 620 °C for 24 hours (DA620) microstructure*

The microstructure of an L-PBF sample directly aged at 620 °C for 24 h is presented in Fig. 12. The BF micrograph of Fig. 12(a) shows that the microstructure consists of dislocation cells and secondary phases (Laves phases, carbides, and oxides) similar to the AP microstructure (Fig. 6). Quantitative analysis (Table 4) shows the particles of Laves phase are 234 ± 46 nm with an interparticle distance of 399 ± 96 nm –statistically unchanged from AP. Interestingly, a high dislocation density ($1.4 \pm 0.4 \times 10^{14}$ m$^{-2}$) similar to the that in the interior cells of the AP condition is observed. This heat treatment was not sufficient to promote recovery of lattice dislocations. Additionally, statistically significant changes in dislocation cell sizes (640 ± 110 nm) and the dislocation entanglements are not observed.

Unlike the AP condition, monolithic precipitates are observed to have formed with average lengths of 8 ± 2 nm. The FFT pattern (inset in Fig. 12(b)) indicates that the nanoprecipitates are γ″, which is expected at this aging temperature. Moreover, Nb elemental mapping (Fig. 12(d)) shows that there is still Nb segregation along the cell boundaries, as in Fig. 6(e), indicating that this time and temperature of aging is insufficient to homogenize Nb throughout the matrix. Conversely, Ti, which is segregated in the AP condition (Fig. 6(f)), is homogenously redistributed using this aging time and temperature.

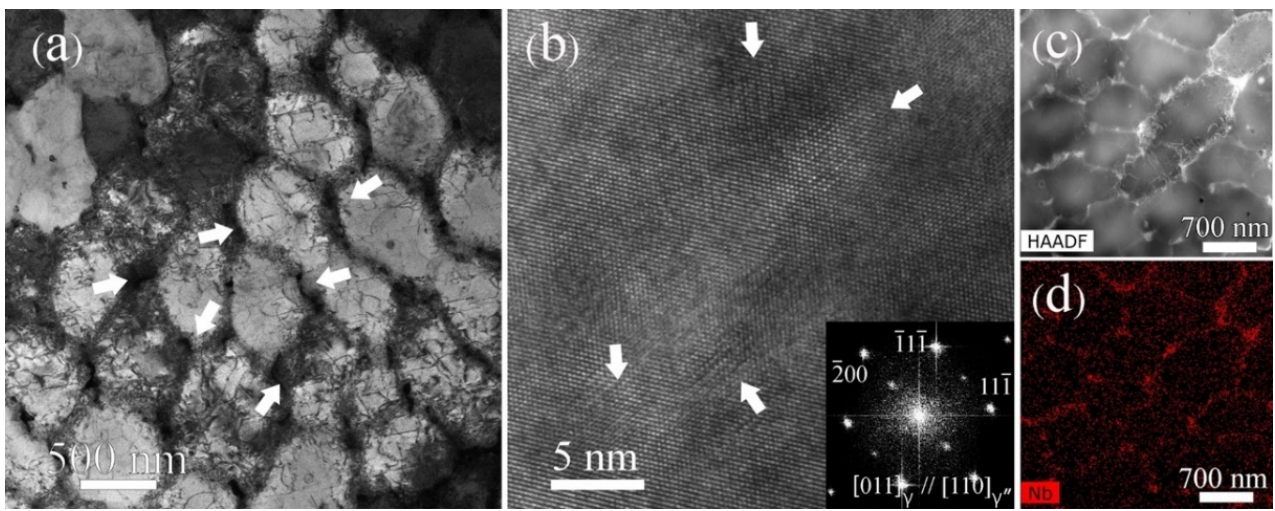

Fig. 12. (a) Conventional BF micrograph after 620 °C for 24 h (DA620) showing secondary particles i.e. Laves phases (indicated by arrows), dislocation cell network and high dislocation density inside cells retained from as-printed condition. (b) High-resolution TEM micrograph taken along $[011]_\gamma$ // $[110]_{\gamma''}$ showing formation of $\gamma''$ nanoprecipitates throughout the $\gamma$ matrix (indicated by white arrows). (c-d) HAADF-STEM micrograph and corresponding Nb elemental map showing remnant Nb segregation along the cell boundaries.

*3.2.6. The direct aged at 720°C for 24 hours (DA720) microstructure*

Much like the 620 °C for 24 h and AP conditions, quantitative analysis (Table 4) of images such as the BF micrograph of the 720 °C for 24 h condition in Fig. 13(a) shows that the microstructure consists of 620 ± 110 nm dislocation cells with cell boundaries consisting of an entangled network of dislocations (Fig. 6), confirming the thermal stability of the dislocation cell structure at these times and temperatures [12,13]. Laves phase particles have an average size of 240 ± 58 nm with an interparticle distance of 380 ±105 nm, also similar to the AP and DA620 conditions. However, the dislocation density within the interior of the cells has decreased to 2.4 ± 0.6 × $10^{13}$ m$^{-2}$, which is nearly an order of magnitude less than the AP condition and approximately 1/3 less than the solution annealed (SA980) condition. This result indicates that unlike 620 °C, the 720 °C aging temperature is sufficient to promote lattice dislocation annihilation and the thermal recovery. Interestingly, aging for 24 h at this lower temperature leads to more dislocation recovery than annealing for 1 h at 980 °C. Some of the remaining lattice dislocations are indicated with white arrows in Fig. 13(a).

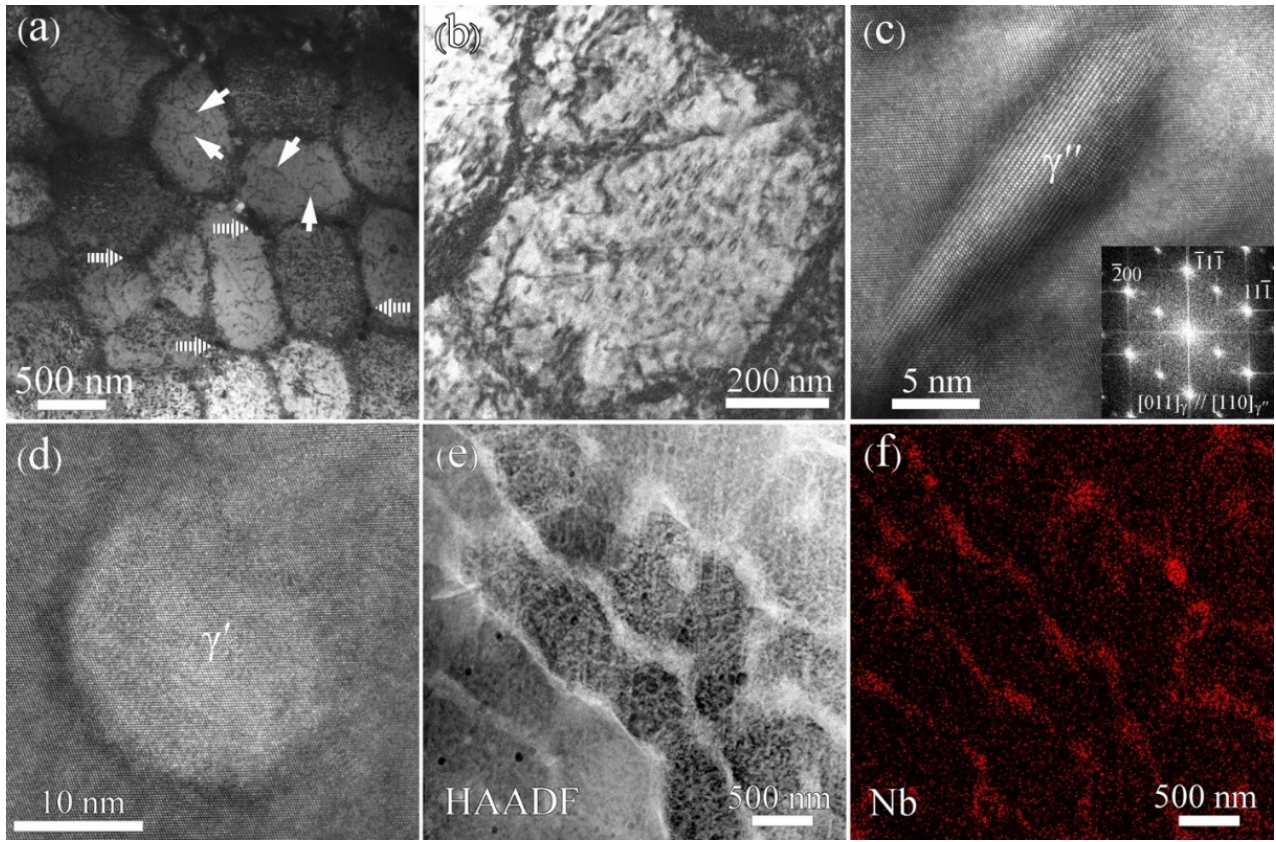

Fig. 13. (a) Conventional BF micrograph of AM-IN718 after direct aging at 720°C, 24h (DA720) showing secondary particles i.e. Laves phases (indicated with striped arrows), dislocation cell network and few dislocations (indicated with white arrows) inside the cells retained from as-printed condition. (b) High magnification BF micrograph showing presence of coprecipitates. High-resolution TEM micrograph taken along $[011]_\gamma$ showing formation of (c) $\gamma''$ and (d) $\gamma'$ nanoprecipitates throughout the γ matrix. (e-f) HAADF-STEM micrograph and corresponding Nb elemental map showing remnant Nb segregation along the cell boundaries.

Additionally, the high magnification BF micrograph in Fig. 13(b) shows the formation of dense precipitate networks. Quantitative analysis (Table 4) of HRTEM images including Fig. 13 (c) and (d) also confirm the existence of monolithic $\gamma''$ (31 ± 8 nm) and $\gamma'$ (21 ± 7 nm), respectively, in addition to γ″/γ′ coprecipitates (15 ± 2 nm) within the interiors of the cells. This nanoprecipitate structure is most similar to the SHT-2 condition. Interestingly, a few larger monolithic γ″ precipitates (210 ± 50 nm) were also observed at the cell boundaries, like those observed of SHT-1, indicating that this DA720 treatment is sufficient to initiate heterogeneous γ′ and γ″ nucleation from the dislocation entanglements. Moreover, Nb elemental mapping (Fig. 13(e-f)) of the DA720 condition shows that there is still Nb segregation along the cell boundaries even after aging at 720 °C for 24 h; therefore, this aging temperature is also insufficient to dissolve Laves particles or to diffuse the segregated Nb back into the matrix. However, similar to the DA620 condition (Fig. 12(d)), Ti solute present in the AP condition is homogenized.

*3.2.7. The solution anneal at 1020 ºC for 0.25 h, age at 720 ºC for 24 h (SA1020+A720) microstructure*

Quantitative analysis of images including the BF micrograph of Fig. 14(a) indicate a microstructure consisting of dislocation cells that are 660 ± 160 nm—statistically equivalent in size to all other aforementioned conditions, though the cell boundaries are much sharper and there is no longer a strong "entanglement" nature to the dislocations that define the cells. The lattice dislocations in the cell interior have been annihilated and recovered (Fig. 14(a)). Moreover, in the BF micrograph of Fig. 14(b), it is apparent that the cell boundaries are pinned by MC carbide and/or oxide nanoparticles through Zener pinning effects [76], which inhibits cell boundary movement at high temperatures. EBSD analysis (not shown) determined the grain size and crystallographic texture to be statistically equivalent for the SA1020+A720 sample relative to the AP condition shown in Fig. 4(b), as expected given established knowledge of times and temperatures required for recrystallization of L-PBF AM-IN718 [12,13].

Dense, homogeneously distributed nanoprecipitation is revealed by diffraction contrast in the Fig. 14(a-b) BF micrographs. Quantification (Table 4) of HRTEM micrographs including Fig. 14(c-d) determine the nanoprecipitation to consist of monolithic γ″ (29 ± 7 nm) and γ′ (23 ± 4 nm) precipitates as well as γ′/γ″ coprecipitation (18 ± 3 nm) separated by 16 ± 5 nm regions of γ matrix phase. In addition, STEM-EDX revealed homogenization of Nb and Ti segregation, similar to that observed for the SA980 condition. Laves phase and δ precipitates were suppressed (Fig. 14(a-b)), as expected from the higher temperature solution anneal.

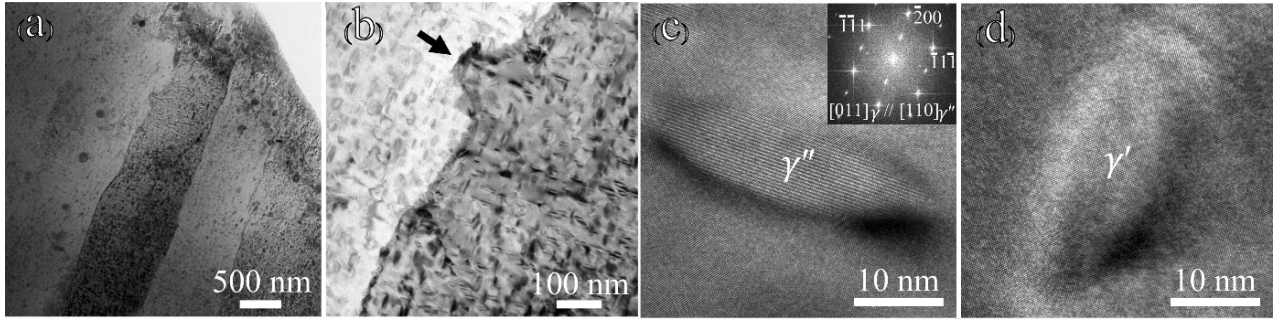

Fig. 14. (a) Conventional BF micrograph of AM-IN718 in SA1020+A720 condition showing dislocation cell network with diffraction contrast from dense nanoprecipitation. (b) High magnification BF micrograph showing presence of nanoprecipitation and Zener pinning of the intercellular boundary by MC carbide particle. High-resolution TEM micrograph taken along $[011]_\gamma$ showing formation of (c) γ″ and (d) γ′ nanoprecipitates throughout the γ matrix.

## 4. Discussion

**4.1 Solute segregation and misorientation periodicities dictate dislocation cell structures**

As shown by Staker and Holt [77] and rearranged by Kocks and Mecking [78], many fcc metals subjected to deformation form dislocation cell substructures that follow the following empirical relationship between dislocation cell diameter, $d$, and bulk dislocation density, $\rho$:

$$\sqrt{\frac{\pi}{4}d^2} = \frac{10}{\sqrt{2\rho}} \quad \text{or} \quad d = \frac{10\sqrt{2}}{\sqrt{\pi\rho}} \quad \text{or} \quad \rho = \frac{200}{\pi d^2} \tag{1}$$

With $d = 620 \pm 120$ nm and $\rho = 1.6 \pm 0.8 \times 10^{14}$ m$^{-2}$, the AP microstructure fits this relationship well. Predicted values fall within one standard deviation: a cell size of 620 nm indicates a dislocation density of $\rho = 1.66 \times 10^{14}$ m$^{-2}$, and conversely, a dislocation density of $\rho = 1.6 \times 10^{14}$ m$^{-2}$ implies a cell size of 631 nm. While the dislocations along cell walls in the AP material were too entangled to include them in dislocation density measurements, the cell walls themselves constitute very low volume fraction within the material, especially considering their columnar geometry (Fig. 6(b)). Hence, even though they are more densely populated with dislocations than the interiors, they have very little influence on the average bulk dislocation density according to the volume-averaging scheme summarized by Kocks and Mecking [78]. For example, if they occupy 5% of the material volume (based on quantification of BF images for AP condition including Fig. 6(a) and (b)) and are populated 10 times more densely with dislocations ($1.6 \pm 0.8 \times 10^{15}$ m$^{-2}$), the average dislocation density would still fall well within one standard deviation of the measurement made from the dislocation density of cell interiors alone.

At first, this agreement between the deformation-induced and the AM-induced dislocation cell parameters may seem coincidental – after all, one structure is formed by mechanical deformation, while the other is a solidification structure driven by solidification morphology, solute atom rejection and thermal stresses. However, the mathematical theory formulated by Holt follows the same rationale used by Cahn and Hilliard to predict periodic fluctuations of solute atom clusters during spinodal decomposition of a supersaturated solution [79]. The prediction that periodic fluctuations in dislocation density are more stable than uniform arrays of dislocations is independent of how the dislocations form.

Moreover, the exact mechanism of dislocations generation during AM is still subject of discussion in literature [80,81]; however, our results in this work are consistent with the hypothesis that the formation of low-energy dislocation cell structures, even after rapid solidification, indicates that significant post-deposition annealing occurs by heat transfer to the already-built regions of the samples from the newly-deposited regions. The size scale of the dislocation cell structure is dictated by the dendrite geometries at the time of solidification, via the rejection of solute atoms to the tip of dendrites [20]. Some geometrically necessary dislocations form at the cell boundaries to accommodate misorientations between dendrite arms at solidification. Further entanglements of dislocations form at the cell walls from thermal stresses upon subsequent deposition passes, tougher with pinning of lattice dislocations by the segregated solute atoms [82]. Furthermore, the dislocation content of the cell interiors is geometrically necessary to stabilize the cellular periodicity of the dislocation entanglements, according to Holt's theory. This result shows that established theory for the recovery of deformation induced dislocation substructures is applicable to AM - induced dislocation substructures.

**4.2 Dislocation cells enhance the yield strength of as processed AM-IN718**

As reviewed in Section 1.1, it is established that dislocation structures inherent to some L-PBF AM alloys can enhance mechanical properties [1,2]. Changes in yield strength can be correlated to the microstructure origins of strengthening mechanisms that operate in AM materials. The Hall–Petch relation is one strengthening model for crystalline materials, and it states that strength is inversely proportional to the square root of grain size associated with HAGBs [19,79]. In first reporting strengthening effects of the dislocation cells within AM materials, Wang et al. proposed that the size of the dislocation cells could be used in place of the size of the grains. Recently, Zhu et al. [3] found that such an approximation drastically overestimates the strength of a L-PBF CoCrFeNiMn high-entropy alloy; at 1278 MPa, it also drastically overestimates the strength of the AP AM-IN718. Instead, they turned to the classical strain-hardening model to account for strengthening from both HAGBs and dislocation densities present in an L-PBF CoCrFeNiMn high-entropy alloy in the as-printed condition:

$$\sigma_y = \sigma_0 + k/\sqrt{d} + \alpha M G b \sqrt{\rho}, \qquad (2)$$

where the first two terms are the classical Hall-Petch relationship and the last term accounts for dislocation Taylor hardening. The applicability of this model to AM metals and the interpretation of calculating $\rho$ using analysis of the cellular dislocation structures is supported by the discussion in Section 4.1.1. Using established values for IN718: friction stress $\sigma_0$ (325 MPa) [83], the Hall–Petch coefficient k (750 MPa μm$^{1/2}$) [84], shear modulus G (80 GPa) [85], Burgers vector b (0.2539 nm) [56], α (0.35) [86], together with measured values of the average grain width from HAGBs d (25.2 μm), dislocation density $\rho$ (1.6 ± 0.8 × 10$^{14}$ m$^{-2}$), and the Taylor factor assessed of the EBSD analysis, M (3.3), the resulting yield strength calculation is 771 MPa.

Furthermore, the cells and grains are columnar, not equiaxed, and thus the size-scale (d) for the Hall-Petch calculation depends on the orientation of the slip systems relative to the columnar grains. For example, in an extreme assumption, slip may be activated such that dislocations only propagate along the length of the columnar grains (Figure 4), This leads to a yield strength of 694 MPa when using d = 108.4 µm in Eq. 2 (all other parameters the same). Given the calculated yield strength can be assumed to be bound by the width and length of the columnar grains, the yield strength estimation is in good agreement with the experimentally observed value of 760 MPa (Table 3, Fig. 3), which is as good as should be expected of this analytical theory [78]. This is especially true considering that this analysis ignores Nb and Ti solutes strengthening, Laves phase strengthening, and dispersion strengthening from oxides and carbides. Nonetheless, this result shows the importance of the dislocation cells as an operative strengthening mechanism since its contribution (the last term in Eq. 2) accounts for a 63% increase in strength, or 39% of the total strength of the AP AM-IN718 material relative to considering GB Hall-Petch relationship alone, which predicts a yield strength of 474 MPa using the 25.2 µm grain size.

### 4.3 Dislocation cells enhance the elongation to failure of AM-IN718

Liu et al. [2] showed that the AM dislocation cells impede dislocation motion, resulting in dislocation storage that leads to higher yield strength, strain hardening, and elongation to failure. They reported that stable plastic flow in AM 316L stainless steel was achieved by maintaining the dislocation network in the microstructure. In the present case of IN718 produced by L-PBF, MC carbide and oxide particles act as pinning sites, stabilizing the dislocation cell structure during deformation, leading to enhanced strain hardening. Note that the carbides in this material are nanosized, hence they most likely do not promote brittle failure since they have a low volume fraction and are well dispersed. Therefore, in the AM-IN718 materials, the motion of dislocations within the cells is hindered by the dislocation cell boundaries (Fig. 6 (a) and (b)), similar to the result shown by Yoo et al. in examining solution heat treated AM-IN718 after 1000 low-cycle tension-tension fatigue cycles [20]. By increasing the strain, slip can transfer across the cells, which leads to the increase of the strength without sacrificing the elongation. As such, the dislocation network has a critical role in the improvement of mechanical properties, and its stability during plastic deformation is of great importance in the enhancement of elongation.

**4.4 Coprecipitates in heat treated AM-IN718**

Contrary to previous reports [87], coprecipitates are present in AM-IN718 after aging, be it direct aging or aging of solutionized materials (Table 4). Detor et al. [31] has shown that coprecipitation can exist over a broad range of (Al+Ti)/Nb ratios. As demonstrated in this work, HR HAADF STEM and high-spatial resolution STEM-EDX were beneficial in definitively confirming the existence of coprecipitation. Using HRTEM, identification of coprecipitates is conducted by identifying the characteristic, Al-rich planar interface (similar to that shown in Fig. 11(b)) and surrounding strain field. Since these techniques were not previously implemented when evaluating AM-IN718, it is plausible the existence of coprecipitates may simply have been overlooked, as the present detailed analysis by high resolution EDS coupled with HRTEM utilized in this study has definitively revealed coprecipitate structures.

In addition, when evaluating the sizes of the different $\gamma''$ and $\gamma'$ precipitate morphologies across the different aging treatments, the monolithic $\gamma''$ are always the largest precipitate, while coprecipitates are the smallest (Table 4, Figs. 6–8). This finding suggests that the coprecipitates coarsen at a slower rate than their individual monolithic phases, which is consistent with previous studies of their kinetics in wrought processes [23,31,32,75]. As discussed in section 3.2.3, optimizing the formation of coprecipitates in AM-IN718 should improve the performance at higher operating temperatures.

**4.5 Microstructures and properties differ in choosing to heat treat individual parts vs. entire builds**

Solution annealing effectively homogenizes chemical segregation, dissolves the Laves phase, and significantly reduces dislocation density (Table 4) on both cell boundaries and cells interiors, collectively resulting in a 23% decrease in YS with a corresponding 43% increase in elongation to failure of the SA980 sample relative to the AP sample. Considering the industry SHTs, as noted in Section 3.2.4, it is found that removing a sample from a build plate after heat treatment results in a loss of cellular arrays of dislocation entanglements in favor of coarse $\gamma''$ and $\gamma'$ precipitates (Fig. 9). This effect is explained by the slower cooling rates achievable due to the large thermal mass when the test pieces remain attached to the substrate during heat treatment.

A higher yield strength in the SHT-1 sample is achieved relative to the SA980 sample through precipitation strengthening (Table 4); a yield strength that exceeds the minimum strength requirement specified in AMS5662. But although SHT-1 AM-IN718 can achieve the current industry standards for wrought IN718 parts, performance of AM-IN718 processed under these conditions is still suboptimal and considerations must be made during post processing to design better heat treatments for AM-IN718. Clearly, these over-coarsened precipitates are undesirable, as they result in a yield strength reduction of ~ 100 MPa and a 1% less elongation relative to SHT-

2. Recovery of the dislocation cell structure reduces both strength and elongation, emphasizing the benefits of keeping the dislocation entanglements at the cell boundaries. Furthermore, the nanoprecipitates are much coarser within the cell interiors. Most notably at ~ 65 × 25 nm, the γ″ precipitates are more than 3 times bigger. These changes are a result of protracted transformation kinetics during slower cooling rates between heat treatment steps. These conditions are caused by radiation and heat conduction into the sample from the build plate and surrounding parts – especially the cooling from the solution anneal. The direct aging studies provide further support for this conclusion; only sparse, coarse, heterogeneous precipitation is observed in DA720 samples, indicating that 720 °C is likely the lower-bound to heterogeneous precipitation. Thus, it is recommended that in heat treating AM-IN718, parts spend as little time as possible at temperatures between the solution anneal temperature and 720 °C.

## 4.6 The coexistence of dislocation cells and fine, dense nanoprecipitates leads to stronger AM materials

Extended-duration, direct-aging heat treatments (without first solution annealing) provide insight into process–structure–property relationships in AM IN718. Based on microstructural characteristics, DA620 samples retain higher dislocation densities within the cell substructure, while the DA720 samples show larger and denser nanoprecipitate morphologies (Table 4). Despite some dislocation content reductions relative to AP conditions, both samples preserve the dense dislocation cell structure. While dislocation cell structures remained relatively constant, γ″ precipitate strengthening dominates dislocation, γ′, and coprecipitate strengthening mechanisms; the strength increases by ~ 50% due solely to minimal γ″ nanoprecipitation in the DA620 sample relative to the AP sample. The DA720 sample also shows an additional 20% rise in yield strength relative to the AP sample, verifying that the γ″ and coprecipitates are critical to achieving maximum strengths. And while both samples demonstrate that precipitation has a very strong influence, dislocation cells enable AM-IN718 to be stronger than optimally processed cast and wrought IN718. In spite of its sub-optimal aging, DA620 achieved comparable strength and superior elongation to wrought, SHT–processed IN718. In contrast, DA720 AM-IN718 achieved greater strengths, but lower elongation. It appears that the strength-elongation paradigm for wrought-SHT material cannot be overcome through direct-aging alone, even absent δ phase precipitation. Laves phase dissolution and/or homogenization of the Ti and Nb solute segregation will be necessary to realize the full potential benefits of the AM-induced dislocation cells in AM-IN718.

## 5. Conclusions

Considering the foundational work reviewed in Section 1, together with these new systematic studies of the individual SHT heat treatment steps and direct aging results, the following process-structure-property understanding is gleaned that can be used to engineer better heat treatments:

- δ phase precipitation is not necessary for strength (this study) or creep performance [4] if AM-induced dislocation cells are retained. Dislocation cell structure and nano-sized oxide and carbide particles stabilize interfaces, mitigating grain growth. Hence, δ phase is detrimental. Perform solution anneals of AM-IN718 at temperatures above 1010 °C (δ-solvus) to eliminate δ phase.

- AM-induced dislocation cells enhance strength and elongation (this study) as well as creep performances [4] of L-PBF AM-IN718 relative to that possible through wrought microstructure engineering; if such enhancements are desired, anneal at temperatures below 1100 °C, where dislocation cells are stabilized against recrystallization through Zener pinning provided by nanoscale oxide and carbide particles (this work and [12,13]);

- While solution anneal temperatures above 1100 °C will recrystallize AM-induced defect structures, they will also coarsen grains as well as carbide and oxide particles [12,13]; hence, "re-setting" the AM-induced microstructures with such recrystallization anneals mitigates the possibility of using AM-induced dislocation cell structures to engineer enhanced properties.

- Laves phase is largely undesirable; solution anneals before aging are still desired to dissolve the Laves particles that form during AM.

- Dense, homogenous networks of $\gamma''$, $\gamma'$, and $\gamma''/\gamma'$ coprecipitates are desired; aging after solution anneal at times and temperatures that promote all three nanoprecipitate morphologies is recommended; however, 2-step aging does not appear necessary. Optimal nanoprecipitate structures even of the $\gamma''$ phase, seem to form after 720 °C aging.

- Heterogenous nucleation of $\gamma''$ and $\gamma'$ precipitates can occur when cooling too slowly from solution annealing treatments. Cooling from the solution anneal temperature to temperatures below 720 °C as quickly as possible is recommended.

The results from examining the heat treatments being used by industry today for AM-IN718 show that the material can meet existing specifications. The results from the newly designed solution anneal at 1020 °C for 15 minutes, water quenching, then aging at 720 °C for 24 h, air cooling (SA1020+A720) heat treatment applied to AM-IN718 samples in this study provide verification that process-structure-property knowledge generated in recent years translates to an ability to design heat treatments that improve both strength and elongation beyond what is observed of traditionally manufactured wrought, SHT IN718. This heat treatment has three differences from the heat treatment reported to exhibit the best-known creep performance of L-PBF AM-IN718 and displays better creep performance than cast and wrought by nearly one decade (solution annealed at 1000 °C for 1 h, water quench, age at 720 °C for 8 h, furnace cool, age at 620 °C for 8 h, furnace cool) [4]:

1) The solution anneal temperature chosen is slightly higher to ensure the anneal is carried out above the δ-solvus temperature;

2) The solution anneal time is lower. It is shown that 15 minutes is sufficient to dissolve Laves phase and homogenize Nb and Ti solute segregation. While quantitative analysis of the dislocation entanglements is challenging, using as short a time as possible would preserve as much of the cellular structure as possible. However, there is no noticeable difference in comparing the cellular structures between the Pröbstle et al. material and the material from this study.

3) A single step age at 720 °C for 24 h is used.

Altogether, the best comparison of the SA1020 + A720 microstructure in this work is very similar to the superior creep microstructure in the work by Pröbstle et al. They concluded that their nanoprecipitate structure is suboptimal sized (< 30 nm) [4]. However, they do not appear to use HRTEM to quantify their size, but rather use dark field diffraction conditions, which are not always sufficient to precisely measure precipitates at these scales. Furthermore, coprecipitation was not reported.

Finally, despite the fact that the materials reported in this work contain some subsurface porosity and microcracking within 100 μm of the surface [88], the as-printed surface finish minimally impacted the mechanical performance of the material; confirming the suggestion of Kantzos et al. [89], who concluded that varying amounts of porosity and microcracking in heat treated L-PBF AM-IN718 has only minor effects on the observed tensile properties (i.e. YS, UTS, % elongation). This conclusion is verified in comparing the results of SA1020 + A720 to SA1020 + A720 + M in the present work (Fig. 3).

**Conflict of interest**

No Conflict of interest

**Acknowledgments**

We gratefully acknowledge support from the Department of Defense, Office of Economic Adjustment (grant no. ST1605-17-02) and the Colorado Office of Economic Development & International Trade (grant no. CTGG1 2016-2166). MJM acknowledges the support of NSF-DMREF program (grant no. 1534826) and the Center for Electron Microscopy and Analysis (CEMAS). Robert Hayes of Metals Technology, Inc. is acknowledged for supplying the wrought-condition, standard heat treatment stress-strain data. Drs. Ronald D. Noebe and Dereck F. Johnson of NASA Glenn Research Center are acknowledged for providing the ICP analyses of IN718 powders and printed parts, presented in Table 1. Dialogue with Profs. Ting Zhu and David McDowell of Georgia Institute of Technology helped develop the discussion given in Section 4.1.

Site specific control of crystallographic grain orientation through electron beam additive manufacturing, Mater. Sci. Technol. 31 (2015) 931–938. doi:10.1179/1743284714Y.0000000734.

[50] N. Raghavan, S. Simunovic, R. Dehoff, A. Plotkowski, J. Turner, M. Kirka, S. Babu, Localized melt-scan strategy for site specific control of grain size and primary dendrite arm spacing in electron beam additive manufacturing, Acta Mater. 140 (2017) 375–387. doi:10.1016/j.actamat.2017.08.038.

[51] K. Yuan, W. Guo, P. Li, J. Wang, Y. Su, X. Lin, Y. Li, Influence of process parameters and heat treatments on the microstructures and dynamic mechanical behaviors of Inconel 718 superalloy manufactured by laser metal deposition, Mater. Sci. Eng. A. 721 (2018) 215–225. doi:10.1016/j.msea.2018.02.014.

[52] S. Kou, Welding Metallurgy, 2nd ed., John Wiley & Sons, Hoboken, NJ, 2003.

[53] Thermo-Calc Software, (2018). http://www.thermocalc.com/products-services/software/thermo-calc/ (accessed December 12, 2018).

[54] W.J. Sames, K.A. Unocic, R.R. Dehoff, T. Lolla, S.S. Babu, Thermal effects on microstructural heterogeneity of Inconel 718 materials fabricated by electron beam melting, J. Mater. Res. 29 (2014) 1920–1930. doi:10.1557/jmr.2014.140.

[55] F. Bayerlein, F. Bodensteiner, C. Zeller, M. Hofmann, M.F. Zaeh, Transient Development of Residual Stresses in Laser Beam Melting - A Neutron Diffraction Study, Addit. Manuf. 24 (2018) 587–594. doi:10.1016/J.ADDMA.2018.10.024.

[56] K.N. Amato, S.M. Gaytan, L.E. Murr, E. Martinez, P.W. Shindo, J. Hernandez, S. Collins,